\documentclass[10pt]{article}    
 \usepackage{graphicx}
\begin{document}

\title{The MIp Toolset: an efficient algorithm for calculating Mutual Information in protein alignments}

\author{Russell J Dickson
      and
         Gregory B Gloor\thanks{Corresponding Author: ggloor@uwo.ca}\\
         Department of Biochemistry, University of Western Ontario,\\ 
         London, ON, Canada
      }
\date{}
\maketitle

\begin{abstract}
        
Background: Coevolution within a protein family is often predicted using statistics that measure the degree of covariation between positions in the protein sequence. Mutual Information is a measure of dependence between two random variables that has been used extensively to predict intra-protein coevolution.
      
Results: Here we provide an algorithm for the efficient calculation of Mutual Information within a protein family. The algorithm uses linked lists which are directly accessed by a pointer array. The linked list allows efficient storage of sparse count data caused by protein conservation. The direct access array of pointers prevents the linked list from being traversed each time it is modified.

Conclusions: This algorithm is implemented in the software MIpToolset, but could also be easily implemented in other Mutual Information based standalone software or web servers. The current implementation in the MIpToolset has been critical in large-scale protein family analysis and real-time coevolution calculations during alignment editing and curation.

The MIpToolset is available at:

https://sourceforge.net/projects/miptoolset/

\end{abstract}

\section*{Introduction}

The identification and analysis of covarying positions in a protein family gives important insights into that family's evolutionary history and provides information about sites that are important for function and structural stability as it is believed that covariation implies coevolution \cite{Atchley:2000,Tillier:2003p171,Gloor:2005vu,Travers:2007}. Coevolutionary analysis of protein families is important because it potentially provides a direct link between primary sequence, in the form of multiple sequence alignments, and structure/function predictions. Covariation between positions in a protein family is assumed to derive from phylogenetic, structural, functional, interaction, and stochastic signals \cite{Atchley:2000}. Decomposing this signal is difficult because the phylogenetic and stochastic signal can overwhelm the structural and functional signal \cite{Martin:2005}. Furthermore, alignment errors have been shown to produce misleading erroneous signal \cite{Dickson:2010p237}.

One of the most popular methods for quantifying covariation in proteins is Mutual Information ($MI$). There are many coevolution prediction methods which are derived from MI \cite{Gloor:2005vu,Dunn:2008,Little:2009,Buslje:2009ct,Dickson:2010p237}. As well, there are many web-based servers which will calculate Mutual Information from a submitted protein alignment \cite{Kozma:uj,Chakraborty:2012cb,Yip:2008,GouveiaOliveira:2009eb}. Despite its simple formulation, calculation of MI is computationally demanding, largely because it must be calculated for all pairs of positions in the alignment, meaning it scales $n^2$ relative to the length of the alignment. Further, calculating inter-protein coevolution requires concatenated alignments which increases the effective number of pairs of positions.

Herein we describe an algorithm for calculating MI in protein alignments with high efficiency. This algorithm allows for database-wide analysis \cite{Dickson:2010p237} and real-time calculation of covariation during alignment curation \cite{Dickson:2012jx}. This algorithm is included as part of the MIpToolset.

\section*{Algorithm}
  \subsection*{Mutual Information}

The calculation and formulation of Mutual Information is described in detail in \cite{Martin:2005}; it is outlined here to provide necessary background to understand the optimizations of the MIpToolset algorithm. 

Mutual Information measures the degree of covariation between two random variables (in our case, protein alignment positions $X$ and $Y$) using the Information Theoretic quantity Entropy ($H$). 

\begin{equation}\label{MI}
MI_{x,y} = H_x+ H_y - H_{x,y}
\end{equation}

Information Entropy ($H$) can be understood as the measure of uncertainty of the identity of the amino acid at some position $x$. As shown in equation \ref{entropy}, the Entropy ($H$) for position $x$ is calculated using the probability of each of the 20 amino acids appearing at that position. Since the actual probabilities are unknown, the amino acid frequencies in the input alignment are used to approximate these values.

\begin{equation}\label{entropy}
H_x= -\sum_{i=1}^{20} p(x_i)\log_{20} p(x_i)
\end{equation}

The $MI$ between positions $X$ and $Y$ is the sum of the Entropy of each position minus the "joint Entropy" between them. The enumeration of joint entropy is the rate-limiting step of Mutual Information calculations. Joint Entropy is calculated similarly to Entropy, but it involves the calculation of probability of all pairs of amino acids that occur between position $x$ and position $y$ (Equation \ref{jointentropy}).

\begin{equation}\label{jointentropy}
H_{x,y}= -\sum_{i=1}^{20}\sum_{j=1}^{20} p(x_i, y_j)\log_{20} p(x_i, y_j)
\end{equation}

The na\"{\i}ve calculation of joint entropy is inefficient because it involves populating a 20 x 20 matrix for every pair of amino acids found for every pair of positions. This is a 400-entry matrix for $n^2$ positions. This approach, while easy to implement, uses an unnecessary amount of memory as it does not exploit the fact that most positions will be moderately conserved and, thus, most positions will have a value of zero in the joint entropy count matrix.

  \subsection*{Storage of sparse matrix in linked list}

\begin{figure}[tbh]
\centerline{\includegraphics{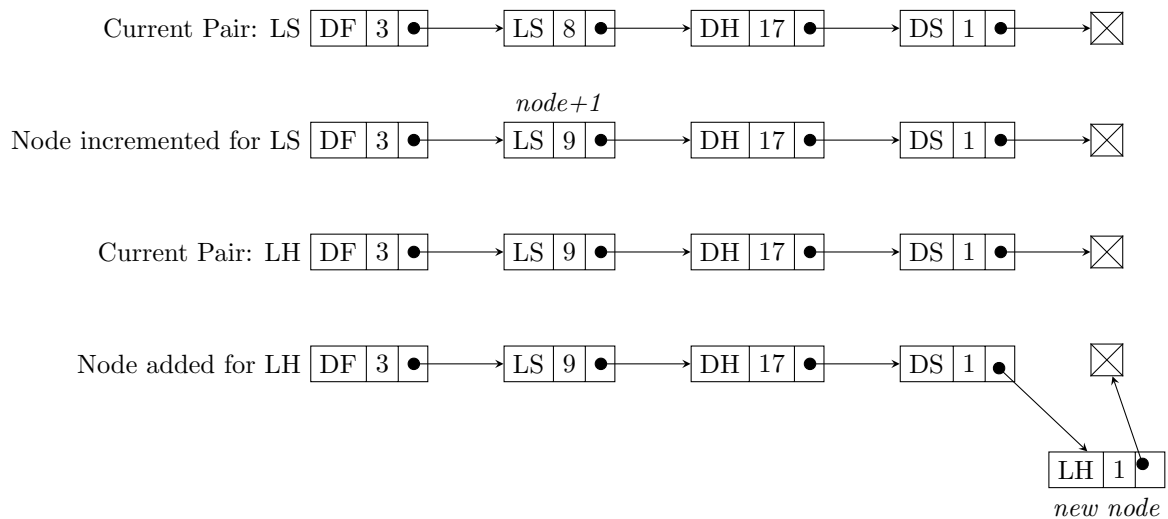}}
\caption{{\bf The linked list storage of amino acid pair counts.} This figure demonstrates how two new amino acid pairs, LS and LH, are added to the growing linked list data structure which stores amino acid pair counts. First, LS is added to the list by incrementing the existing LS node. Second, the pair LH is added to the list by creating a new node labeled LH and adding it to the list with counter set to 1. }
\label{fig1}
\end{figure}

It is worth noting that calculation of $MI$ involves two types of "pairs": Pairs of $positions$, which represent the homologous `columns' in a protein family multiple sequence alignment (MSA), and pairs of $amino~acids$, which are the corresponding entries from a pair of positions within a single sequence. So a pair of positions, might be position 10 and position 45 within a protein sequence; at this pair of positions, there will be many amino acid pairs corresponding to the identity of the amino acids at positions 10 and 45 in the sequence (ie. DF, LS, DH etc.).

A straightforward way to store the counts between positions $x$ and $y$ is to use a linked list data structure (Figure \ref{fig1}). Each node in the linked list stores two values for the calculation of Joint Entropy, the identity of the amino acid pair, and the respective count. Each node also contains a pointer to the next node in the list, or $null$ if the node is the terminal node.

The program iterates over the protein alignment, enumerating the amino acid pairs, just as it would if it were in the na\"{\i}ve implementation. If an entry in the linked list exists for a given amino acid pair, the node's counter is incremented. If no such entry exists a new node is appended to the end of the list for that amino acid pair. This list can be traversed efficiently as these counts will be used for future calculations. This efficient storage makes it possible to efficiently analyze very long alignments.

  \subsection*{Direct access to linked list improves speed}

\begin{figure}[tbh]
\centerline{\includegraphics[width=0.8\textwidth]{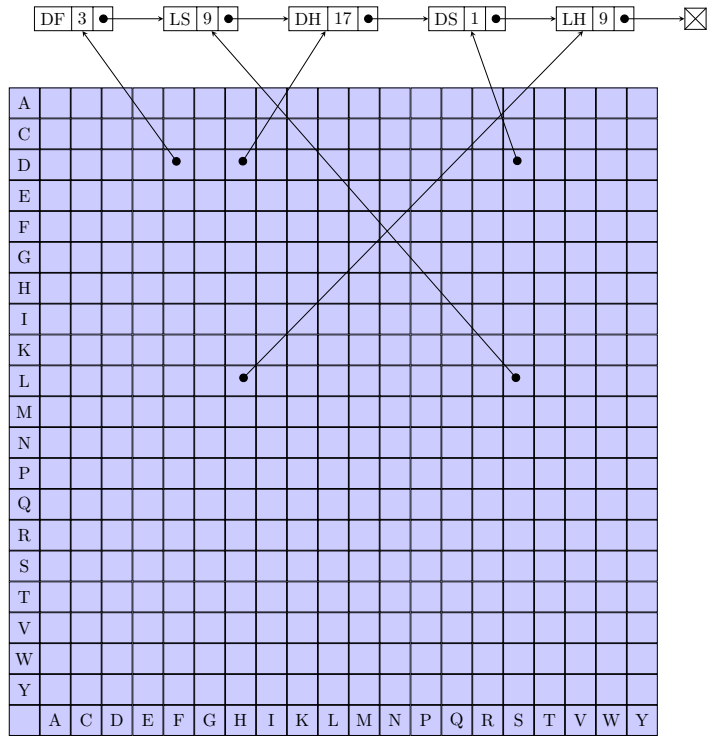}}
\caption{{\bf  Direct-access array of pointers to growing linked list.} This figure demonstrates how the direct-access array provides instant access to any part of the linked list without the need to traverse the list. This array is only allocated once and can be reused for each pair of positions.}
\label{fig2}
\end{figure}

The limitation of the linked list storage method, if a linked list is used on its own, is that the list will need to be traversed each time a node is to be updated or created to check whether that pair exists in the data structure. This challenge can be overcome by using an array of pointers to linked list nodes. The disadvantage of the linked list storage solution is that it lacks "direct access" provided by a two-dimensional array in from the na\"{\i}ve implementation. By combining the two, it is possible to achieve a ``best of both worlds'' solution.

A single "direct-access" 20 x 20 array is created, with the nodes in the array corresponding to the 400 possible amino acid pairs (Figure \ref{fig2}). When an amino acid pair is encountered by the main count enumeration loop, the direct-access array is checked. If the entry for that pair is $null$, then a new linked node is appended to the end of the growing linked list for that pair of positions with a count of 1; next, the entry in the direct-access matrix is set as a pointer to the newly created linked list node.

Conversely, if the entry corresponding to the amino acid pair contains a pointer, the program follows the pointer to the corresponding linked list node and increments the counter by 1. After the two positions have been fully enumerated, all entries in the direct-access array are reset to $null$ and it can be reused. Thus, the direct-access array strategy maintains the advantages of a linked list storage solution without the disadvantage of needing to traverse the list every time, at the trade-off cost of only 400 pointers.

  \subsection*{Integration in the MIpToolset}
  
  This algorithm has been included as part of the MIpToolset, a collection of C- and Perl-based programs which calculate covariation statistics and inter-residue distances from protein alignments and databases. A full description of sequence collection and alignment is available in (Dickson and Gloor, Methods Mol. Biol. 2013 $submitted$). 
  
  In brief, the input to the program is a protein alignment containing more than 150 sequences less than 90\% identical and containing more than 50 ungapped positions. It is recommended that the alignment be manually analyzed by the investigator to ensure the alignment does not contain errors which will lead to false-positive results \cite{Dickson:2010p237,Dickson:2012jx}. For example, the curation tool LoCo \cite{Dickson:2012jx}, based on the alignment viewer Jalview \cite{Waterhouse:2009}, provides a visualization of the likely-misaligned regions of the alignment. The program also optionally accepts a PDB structure corresponding to a sequence in the protein family. This structure is used to generate inter-residue distances which are commonly used to validate coevolution predictions.
  
  The output of the program is a large list of pairs of positions and their corresponding covariation statistics. The MIpToolset presently generates Mutual Information, as well as several more accurate derivations including $MIp$ (and its normalized counterpart $Zp$) \cite{Dunn:2008}, $Zpx$ and $\Delta Zp$ \cite{Dickson:2010p237}. A coevolution network file is also produced which can be visualized using Graphviz \cite{Ellson:2002wf}.

\section*{Conclusions}

It is established that $MI$ by itself is not particularly accurate in predicting coevolving positions because it correlates with Entropy \cite{Martin:2005}, misleading phylogenetic signal \cite{Dunn:2008}, and alignment errors \cite{Dickson:2010p237}. Furthermore, analyzing gaps as the "21st amino acid" causes misleading results which is partially why the aforementioned studies excluded positions containing gaps from the analysis (Dickson et al. submitted). It is possible to overcome some limitations of raw $MI$ by using various corrections to $MI$ \cite{Dunn:2008,Little:2009,Buslje:2009ct,Dickson:2010p237}. Typically these corrections based on an analysis of raw $MI$ values are computationally inexpensive and so heavy optimization is not necessary. Thus the algorithm and software described herein can be used to reduce the time and memory required to calculate most $MI$-derived statistics. 

The MIpToolset has been tested on Unix-like operating systems and is implemented in C for efficiency, with a Perl wrapper for handling input/output issues. The speed and efficiency of the MIpToolset has allowed for efficient database-wide analysis \cite{Dickson:2010p237} and the detection of protein family misalignments using an $MI$-derived method in real-time as the user edits their alignment in the software tool LoCo \cite{Dickson:2012jx}. To our knowledge, this is the fastest implementation of the coevolution statistics $MIp$, $Zp$, $Zpx$, and $\Delta Zp$\cite{Dunn:2008, Dickson:2010p237}.

It is available at: https://sourceforge.net/projects/miptoolset/

\section*{Authors contributions}
RJD designed the algorithm, and wrote the manuscript. GBG designed the project. All authors contributed to the software, and read and approved the final manuscript.

\section*{Acknowledgements}
  The authors wish to thank the students of the 2011 and 2012 Biochemistry 4445a class at UWO for identifying documentation ambiguities and cross-platform issues. Also, thank you to Stan Dunn, Lindi Wahl, David Edgell, Thomas McMurrough, Jean Macklaim, Andrew Fernandes, Ardeshir Goliaei for helpful discussions and testing. Finally thank you to existing users of the MIpToolset for their feedback while using this software while it was in development.

 \bibliographystyle{bmc_article}  
  \bibliography{paperslibrary}      


\begin{thebibliography}{10}
\providecommand{\url}[1]{[#1]}
\providecommand{\urlprefix}{}

\bibitem{Atchley:2000}
Atchley WR, Wollenberg KR, Fitch WM, Terhalle W, Dress AW:
  \textbf{{Correlations among amino acid sites in bHLH protein domains: an
  information theoretic analysis.}} \emph{Mol Biol Evol} 2000,
  \textbf{17}:164--178.

\bibitem{Tillier:2003p171}
Tillier E, Lui T: \textbf{{Using multiple interdependency to separate
  functional from phylogenetic correlations in protein alignments}}.
  \emph{Bioinformatics} 2003, \textbf{19}(6):750--755.

\bibitem{Gloor:2005vu}
Gloor GB, Martin LC, Wahl LM, Dunn SD: \textbf{{Mutual information in protein
  multiple sequence alignments reveals two classes of coevolving positions.}}
  \emph{Biochemistry} 2005, \textbf{44}(19):7156--7165.

\bibitem{Travers:2007}
Travers SAA, Fares MA: \textbf{{Functional coevolutionary networks of the
  Hsp70-Hop-Hsp90 system revealed through computational analyses.}} \emph{Mol
  Biol Evol} 2007, \textbf{24}(4):1032--1044.

\bibitem{Martin:2005}
Martin LC, Gloor GB, Dunn SD, Wahl LM: \textbf{{Using information theory to
  search for co-evolving residues in proteins.}} \emph{Bioinformatics} 2005,
  \textbf{21}(22):4116--4124.

\bibitem{Dickson:2010p237}
Dickson R, Wahl L, Fernandes A, Gloor G: \textbf{{Identifying and seeing beyond
  multiple sequence alignment errors using intra-molecular protein
  covariation}}. \emph{PLoS ONE} 2010, \textbf{5}(6):e11082.

\bibitem{Dunn:2008}
Dunn SD, Wahl LM, Gloor GB: \textbf{{Mutual information without the influence
  of phylogeny or entropy dramatically improves residue contact prediction}}.
  \emph{Bioinformatics} 2008, \textbf{24}(3):333--340.

\bibitem{Little:2009}
Little DY, Chen L: \textbf{{Identification of Coevolving Residues and
  Coevolution Potentials Emphasizing Structure, Bond Formation and Catalytic
  Coordination in Protein Evolution}}. \emph{PLoS ONE} 2009,
  \textbf{4}(3):e4762.

\bibitem{Buslje:2009ct}
Buslje CM, Santos J, Delfino JM, Nielsen M: \textbf{{Correction for phylogeny,
  small number of observations and data redundancy improves the identification
  of coevolving amino acid pairs using mutual information.}}
  \emph{Bioinformatics} 2009, \textbf{25}(9):1125--1131.

\bibitem{Kozma:uj}
Kozma D, Simon I, Tusn{\'a}dy GE: \textbf{{CMWeb: an interactive on-line tool
  for analysing residue-residue contacts and contact prediction methods.}}
  \emph{Nucleic Acids Res} 2012, \textbf{40}(Web Server issue):W329--33.

\bibitem{Chakraborty:2012cb}
Chakraborty A, Mandloi S, Lanczycki CJ, Panchenko AR, Chakrabarti S:
  \textbf{{SPEER-SERVER: a web server for prediction of protein specificity
  determining sites.}} \emph{Nucleic Acids Res} 2012, \textbf{40}(Web Server
  issue):W242--8.

\bibitem{Yip:2008}
Yip KY, Patel P, Kim PM, Engelman DM, McDermott D, Gerstein M: \textbf{{An
  integrated system for studying residue coevolution in proteins.}}
  \emph{Bioinformatics} 2008, \textbf{24}(2):290--292.

\bibitem{GouveiaOliveira:2009eb}
Gouveia-Oliveira R, Roque FS, Wernersson R, Sicheritz-Ponten T, Sackett PW,
  M{\o}lgaard A, Pedersen AG: \textbf{{InterMap3D: predicting and visualizing
  co-evolving protein residues.}} \emph{Bioinformatics} 2009,
  \textbf{25}(15):1963--1965.

\bibitem{Dickson:2012jx}
Dickson RJ, Gloor GB: \textbf{{Protein sequence alignment analysis by local
  covariation: coevolution statistics detect benchmark alignment errors.}}
  \emph{PLoS ONE} 2012, \textbf{7}(6):e37645.

\bibitem{Waterhouse:2009}
Waterhouse AM, Procter JB, Martin DMA, Clamp M, Barton GJ: \textbf{{Jalview
  Version 2--a multiple sequence alignment editor and analysis workbench}}.
  \emph{Bioinformatics} 2009, \textbf{25}(9):1189--1191.

\bibitem{Ellson:2002wf}
Ellson J, Gansner E, Koutsofios L, North SC, Woodhull G:
  \textbf{{Graphviz---open source graph drawing tools}}. \emph{Lecture Notes in
  Computer Science} 2002, :483--484.

\end{thebibliography}

\newcommand{\BMCxmlcomment}[1]{}

\BMCxmlcomment{

<refgrp>

<bibl id="B1">
  <title><p>{Correlations among amino acid sites in bHLH protein domains: an
  information theoretic analysis.}</p></title>
  <aug>
    <au><snm>Atchley</snm><fnm>W R</fnm></au>
    <au><snm>Wollenberg</snm><fnm>K R</fnm></au>
    <au><snm>Fitch</snm><fnm>W M</fnm></au>
    <au><snm>Terhalle</snm><fnm>W</fnm></au>
    <au><snm>Dress</snm><fnm>A W</fnm></au>
  </aug>
  <source>Mol Biol Evol</source>
  <pubdate>2000</pubdate>
  <volume>17</volume>
  <issue>1</issue>
  <fpage>164</fpage>
  <lpage>-178</lpage>
</bibl>

<bibl id="B2">
  <title><p>{Using multiple interdependency to separate functional from
  phylogenetic correlations in protein alignments}</p></title>
  <aug>
    <au><snm>Tillier</snm><fnm>ERM</fnm></au>
    <au><snm>Lui</snm><fnm>TWH</fnm></au>
  </aug>
  <source>Bioinformatics</source>
  <pubdate>2003</pubdate>
  <volume>19</volume>
  <issue>6</issue>
  <fpage>750</fpage>
  <lpage>-755</lpage>
</bibl>

<bibl id="B3">
  <title><p>{Mutual information in protein multiple sequence alignments reveals
  two classes of coevolving positions.}</p></title>
  <aug>
    <au><snm>Gloor</snm><fnm>GB</fnm></au>
    <au><snm>Martin</snm><fnm>LC</fnm></au>
    <au><snm>Wahl</snm><fnm>LM</fnm></au>
    <au><snm>Dunn</snm><fnm>SD</fnm></au>
  </aug>
  <source>Biochemistry</source>
  <pubdate>2005</pubdate>
  <volume>44</volume>
  <issue>19</issue>
  <fpage>7156</fpage>
  <lpage>-7165</lpage>
</bibl>

<bibl id="B4">
  <title><p>{Functional coevolutionary networks of the Hsp70-Hop-Hsp90 system
  revealed through computational analyses.}</p></title>
  <aug>
    <au><snm>Travers</snm><fnm>SAA</fnm></au>
    <au><snm>Fares</snm><fnm>MA</fnm></au>
  </aug>
  <source>Mol Biol Evol</source>
  <pubdate>2007</pubdate>
  <volume>24</volume>
  <issue>4</issue>
  <fpage>1032</fpage>
  <lpage>-1044</lpage>
</bibl>

<bibl id="B5">
  <title><p>{Using information theory to search for co-evolving residues in
  proteins.}</p></title>
  <aug>
    <au><snm>Martin</snm><fnm>L C</fnm></au>
    <au><snm>Gloor</snm><fnm>G B</fnm></au>
    <au><snm>Dunn</snm><fnm>S D</fnm></au>
    <au><snm>Wahl</snm><fnm>L M</fnm></au>
  </aug>
  <source>Bioinformatics</source>
  <pubdate>2005</pubdate>
  <volume>21</volume>
  <issue>22</issue>
  <fpage>4116</fpage>
  <lpage>-4124</lpage>
</bibl>

<bibl id="B6">
  <title><p>{Identifying and seeing beyond multiple sequence alignment errors
  using intra-molecular protein covariation}</p></title>
  <aug>
    <au><snm>Dickson</snm><fnm>RJ</fnm></au>
    <au><snm>Wahl</snm><fnm>LM</fnm></au>
    <au><snm>Fernandes</snm><fnm>AD</fnm></au>
    <au><snm>Gloor</snm><fnm>GB</fnm></au>
  </aug>
  <source>PLoS ONE</source>
  <pubdate>2010</pubdate>
  <volume>5</volume>
  <issue>6</issue>
  <fpage>e11082</fpage>
</bibl>

<bibl id="B7">
  <title><p>{Mutual information without the influence of phylogeny or entropy
  dramatically improves residue contact prediction}</p></title>
  <aug>
    <au><snm>Dunn</snm><fnm>S D</fnm></au>
    <au><snm>Wahl</snm><fnm>L M</fnm></au>
    <au><snm>Gloor</snm><fnm>G B</fnm></au>
  </aug>
  <source>Bioinformatics</source>
  <pubdate>2008</pubdate>
  <volume>24</volume>
  <issue>3</issue>
  <fpage>333</fpage>
  <lpage>-340</lpage>
</bibl>

<bibl id="B8">
  <title><p>{Identification of Coevolving Residues and Coevolution Potentials
  Emphasizing Structure, Bond Formation and Catalytic Coordination in Protein
  Evolution}</p></title>
  <aug>
    <au><snm>Little</snm><fnm>DY</fnm></au>
    <au><snm>Chen</snm><fnm>L</fnm></au>
  </aug>
  <source>PLoS ONE</source>
  <pubdate>2009</pubdate>
  <volume>4</volume>
  <issue>3</issue>
  <fpage>e4762</fpage>
</bibl>

<bibl id="B9">
  <title><p>{Correction for phylogeny, small number of observations and data
  redundancy improves the identification of coevolving amino acid pairs using
  mutual information.}</p></title>
  <aug>
    <au><snm>Buslje</snm><fnm>CM</fnm></au>
    <au><snm>Santos</snm><fnm>J</fnm></au>
    <au><snm>Delfino</snm><fnm>JM</fnm></au>
    <au><snm>Nielsen</snm><fnm>M</fnm></au>
  </aug>
  <source>Bioinformatics</source>
  <pubdate>2009</pubdate>
  <volume>25</volume>
  <issue>9</issue>
  <fpage>1125</fpage>
  <lpage>-1131</lpage>
</bibl>

<bibl id="B10">
  <title><p>{CMWeb: an interactive on-line tool for analysing residue-residue
  contacts and contact prediction methods.}</p></title>
  <aug>
    <au><snm>Kozma</snm><fnm>D</fnm></au>
    <au><snm>Simon</snm><fnm>I</fnm></au>
    <au><snm>Tusn{\'a}dy</snm><fnm>GE</fnm></au>
  </aug>
  <source>Nucleic Acids Res</source>
  <pubdate>2012</pubdate>
  <volume>40</volume>
  <issue>Web Server issue</issue>
  <fpage>W329</fpage>
  <lpage>-33</lpage>
</bibl>

<bibl id="B11">
  <title><p>{SPEER-SERVER: a web server for prediction of protein specificity
  determining sites.}</p></title>
  <aug>
    <au><snm>Chakraborty</snm><fnm>A</fnm></au>
    <au><snm>Mandloi</snm><fnm>S</fnm></au>
    <au><snm>Lanczycki</snm><fnm>CJ</fnm></au>
    <au><snm>Panchenko</snm><fnm>AR</fnm></au>
    <au><snm>Chakrabarti</snm><fnm>S</fnm></au>
  </aug>
  <source>Nucleic Acids Res</source>
  <pubdate>2012</pubdate>
  <volume>40</volume>
  <issue>Web Server issue</issue>
  <fpage>W242</fpage>
  <lpage>-8</lpage>
</bibl>

<bibl id="B12">
  <title><p>{An integrated system for studying residue coevolution in
  proteins.}</p></title>
  <aug>
    <au><snm>Yip</snm><fnm>KY</fnm></au>
    <au><snm>Patel</snm><fnm>P</fnm></au>
    <au><snm>Kim</snm><fnm>PM</fnm></au>
    <au><snm>Engelman</snm><fnm>DM</fnm></au>
    <au><snm>McDermott</snm><fnm>D</fnm></au>
    <au><snm>Gerstein</snm><fnm>M</fnm></au>
  </aug>
  <source>Bioinformatics</source>
  <pubdate>2008</pubdate>
  <volume>24</volume>
  <issue>2</issue>
  <fpage>290</fpage>
  <lpage>-292</lpage>
</bibl>

<bibl id="B13">
  <title><p>{InterMap3D: predicting and visualizing co-evolving protein
  residues.}</p></title>
  <aug>
    <au><snm>Gouveia Oliveira</snm><fnm>R</fnm></au>
    <au><snm>Roque</snm><fnm>FS</fnm></au>
    <au><snm>Wernersson</snm><fnm>R</fnm></au>
    <au><snm>Sicheritz Ponten</snm><fnm>T</fnm></au>
    <au><snm>Sackett</snm><fnm>PW</fnm></au>
    <au><snm>M{\o}lgaard</snm><fnm>A</fnm></au>
    <au><snm>Pedersen</snm><fnm>AG</fnm></au>
  </aug>
  <source>Bioinformatics</source>
  <pubdate>2009</pubdate>
  <volume>25</volume>
  <issue>15</issue>
  <fpage>1963</fpage>
  <lpage>-1965</lpage>
</bibl>

<bibl id="B14">
  <title><p>{Protein sequence alignment analysis by local covariation:
  coevolution statistics detect benchmark alignment errors.}</p></title>
  <aug>
    <au><snm>Dickson</snm><fnm>RJ</fnm></au>
    <au><snm>Gloor</snm><fnm>GB</fnm></au>
  </aug>
  <source>PLoS ONE</source>
  <pubdate>2012</pubdate>
  <volume>7</volume>
  <issue>6</issue>
  <fpage>e37645</fpage>
</bibl>

<bibl id="B15">
  <title><p>{Jalview Version 2--a multiple sequence alignment editor and
  analysis workbench}</p></title>
  <aug>
    <au><snm>Waterhouse</snm><fnm>AM</fnm></au>
    <au><snm>Procter</snm><fnm>JB</fnm></au>
    <au><snm>Martin</snm><fnm>DMA</fnm></au>
    <au><snm>Clamp</snm><fnm>M</fnm></au>
    <au><snm>Barton</snm><fnm>GJ</fnm></au>
  </aug>
  <source>Bioinformatics</source>
  <pubdate>2009</pubdate>
  <volume>25</volume>
  <issue>9</issue>
  <fpage>1189</fpage>
  <lpage>-1191</lpage>
</bibl>

<bibl id="B16">
  <title><p>{Graphviz---open source graph drawing tools}</p></title>
  <aug>
    <au><snm>Ellson</snm><fnm>J</fnm></au>
    <au><snm>Gansner</snm><fnm>E</fnm></au>
    <au><snm>Koutsofios</snm><fnm>L</fnm></au>
    <au><snm>North</snm><fnm>SC</fnm></au>
    <au><snm>Woodhull</snm><fnm>G</fnm></au>
  </aug>
  <source>Lecture Notes in Computer Science</source>
  <pubdate>2002</pubdate>
  <fpage>483</fpage>
  <lpage>-484</lpage>
</bibl>

</refgrp>
} 

\end{document}